\documentstyle[titlepage,fleqn,11pt]{article}
\begin{document}
\begin{titlepage}
\title{Spatiotemporal complexity of the universe at subhorizon scales}
\author{Haret C. Rosu \\
\\
Instituto de F\'{\i}sica de la Universidad de Guanajuato,\\
Apdo Postal E-143, 37150 L\'eon, Gto, M\'exico}
\date{}
{\baselineskip=20pt

\begin{abstract}
This is a short note on the {\em spatiotemporal complexity} of the dynamical
state(s) of the universe at subhorizon scales (up to 300 Mpc).
There are reasons, based mainly on infrared radiative divergences,
to believe that one can encounter a flicker
noise in the time domain, while in the space domain, the
scaling laws are reflected in the
(multi)fractal distribution of galaxies and their clusters. There exist
recent suggestions on a unifying treatment of these two aspects within the
concept of spatiotemporal complexity of dynamical systems driven out of
equilibrium. Spatiotemporal complexity of the subhorizon dynamical state(s)
of the universe is a conceptually nice idea and may lead to progress in our
understanding of the material structures at large scales.

\vskip 2cm


IFUG-26/94 [$\cal H \cal C \cal R$]; Nuovo Cimento B 110, 457-459 (April 1995)

astro-ph/9411112

\vskip 1cm

% draft as at November 27th, 1994

\end{abstract}

    }
\maketitle
\end{titlepage}
\baselineskip=30pt
\parskip=0pt

%%%%%%%%%%%%%%%%%%%%%%%%%%%%%%%%%%%%%%%%%%%%%%%%%%%%%%%%%%%%%%%%%%%%%%%%

The problem of large scale structure of the universe, that is the
understanding of the hierarchical
large-scale clustering of galaxies, is one of the fundamental
topics in modern cosmology \cite{peeb}.

At subhorizon scales of less than about 300 Mpc the structure of the
universe, as reflected, e.g., in the two-point correlation functions of
galaxies, cluster of galaxies, and quasars, shows a common fractal behavior
of the type $r^{-\gamma}$ with $\gamma \sim 1.8$. There exists already an
extensive `(multi-) fractal' literature on detailed aspects of the scaling
properties of the large-scale structure \cite{bor}. Fractals are not the
only candidates in providing a quantitative analysis of the subhorizon
cosmological structure.
Various kinds of percolations \cite{ovru} as well as nonlinear diffusions
either in the same percolation approach or defined through some
nonlinear- noisy differential equations expressing large-scale
turbulent/stochastic behavior \cite{turb} are powerful alternatives.
For example, the above-mentioned result on the $\gamma$ exponent is close
both to the fractal dimension of self- similar percolation clusters
($d_f$= 1.89) and to the fractal dimension of the hull of the diffusion
front ($d_f$= 1.75).

An interesting and far from innoccuous question that one can ask is the
nature of the dynamical state(s) of the Universe [DSU] at subhorizon scales.
It is the goal of this note to provide some general heuristic arguments
concerning the DSU. The arguments are based on a number of interesting
results/claimings scattered in the literature of several distinct topics.
Our point here is to emphasize the connection of the DSU problem with the
general framework of
{\em spatiotemporal complexity} of dynamical systems in
nonequilibrium conditions, such as growth phenomena.
Spatiotemporal complexity
is a concept that applies to a great variety of physical systems displaying
both fractal scaling behavior and (generalized) flicker noise. For
nonequilibrium steady-states entailing the two features the self-organized
criticality (SOC) has been proposed as an explanation \cite{soc}.
Very recently, Maslov, Paczuski, and Bak [MPB] \cite{bak} formally
established the
relationship between spatial fractal behavior and `long-range' temporal
correlations for a broad range of nonequilibrium models (not necessarily of
SOC type). They used the language of avalanche activity, but in fact this
general
result can be obtained by means of fractal renewal processes too as shown,
actually before MPB, by Lowen and Teich [LT] \cite{teic}.
Both avalanche activity and fractal renewal processes can be used in
subhorizon cosmology.
Essentially the result of MPB and LT states that the time correlations in
the local avalanche activity and the spatial activity are different cuts
in the same spatiotemporal fractal. The power spectrum of the generalized
flicker noise found by MPB is of the type $S(f)\sim 1/f^{d_r}$, where
$0\leq d_r \leq 1$ can be expressed in a definite way in terms of the spatial
dimension, the avalanche dimension, and the dynamical exponent \cite{bak}.

Let us pass now to the subhorizon cosmological structures. One would like
to identify somehow spatiotemporal fractals in this case. In the space
domain we already referred to the fractal treatment of the hierarchical
large-scale structure and we have to comment on the cosmic flicker
noise.

There are more than 15 years since Press \cite{p1} has commented on
the possible divergence of the low-frequency fluctuations in the light curves
of quasars. Unfortunately the availability of the astrophysical literature
is quite limited for me at the present time so that I am not aware of the
current situation on this matter.
Therefore I shall turn to another possible line of putting into evidence
some type of cosmic flicker noise. It has to do with the problem of infrared
divergences (IRD) of the massless scalar Green functions in de Sitter space
(or of
minimally coupled scalar fields) \cite{bd}, also known to occur in spatially
flat FRW models with power law expansions \cite{fp}. Papers by Tsamis and
Woodard \cite{tw} on the relaxation of the cosmological constant through
infrared effects in quantum gravity are of relevance to this context too.
To be quoted yet is a
paper of Folacci \cite{f2}, in which he showed that the massless scalar
field theory on de Sitter space is not invariant under the symmetry
group of that spacetime O(1,4). Therefore the infrared divergence is a
real one in contrast to what is happening in the case of the same theory
on the four-dimensional sphere where it is only a gauge artifact. According
to Folacci the same type of results applies for massive scalar fields
for special values of the mass parameter.

The relevance of these IRD results for the subhorizon
spatiotemporal complexity reveals itself when one is invoking the quite
well-known theory of Handel on the quantum infrared origin of the 1/f noise
\cite{hand}. This theory is based on the infrared-type divergence present in
all cross sections, and also in some autocorrelation functions due to
interaction of the current carriers with massless infraquanta whose nature
varies from case to case. Moreover, Handel generalized his approach to
infrared radiative corrections to include the presence of a thermal radiation
background \cite{hand1} and commented on the so-called ``gravidynamic quantum
1/f noise" \cite{hand2}, i.e., the effects of infragravitons in shaping
the distribution of matter in space and time.

Finally, another source of 1/f cosmic noise might be the volume Casimir effect
\cite{sok} but I shall not address this point here.

In conclusion, what I have done in this note was to hint upon the two
aspects required by the spatiotemporal complexity (spatiotemporal fractals)
of the universe and I think that this concept should be seriously considered
by authors dealing with subhorizon scales. The above heuristic comments on the
spatiotemporal complexity of subhorizon organization of the universe might be
useful for further more detailed insight into a problem of much current
interest in astrophysics and cosmology. It would be of interest to determine
the scales at which the universe is trapped into self-organized states, if any.

%%%%%%%%%%%%%%%%%%%%%%%%%%%%%%%%%%%%%%%%%%%%%%%%%%%%%%%%%%%%%%%%%%%%%%%%%%%%%%

\section*{Acknowledgements}

This work was supported in part by the CONACyT Project 4862-E9406.
The author is grateful to Professors R.P. Woodard and P.H. Handel for kindly
providing him with their papers.

%%%%%%%%%%%%%%%%%%%%%%%%%%%%%%%%%%%%%%%%%%%%%%%%%%%%%%%%%%%%%%%%%%%%%%%%%%%%%%%


\begin{thebibliography}{99}
\bibitem{peeb} P.J.E. Peebles, {\em The Large Scale Structure of the Universe}
(Princeton University Press, Princeton, NJ, 1980); {\em Principles of Physical
Cosmology} (Princeton Univ. Press, Princeton, NJ, 1993)
\bibitem{bor} S. Borgani et  al., Phys. Rev. E {\bf 47}, 3879 (1993);
S. Borgani, ``Scaling in the Universe",
Phys. Rept {\bf 251}, 1 (1995), and references therein (available also as 
astro-ph/9404054).
\bibitem{ovru}
B.A. Ovrut, ``Large scale structure and percolation theory", UPR-0521T, 1992,
Invited talk at the XV Int. Warsaw Meeting on Elementary Particle Physics,
Kasimierz, Poland, May 25-29, 1992
\bibitem{turb}
S.N. Gurbatov, A.I. Saichev, and S.F. Shandarin, Mon. Not. R. Astron. Soc.
{\bf 236}, 385 (1989); A. Berera and L.Z. Fang, Phys. Rev. Lett. {\bf 72},
458 (1994)
\bibitem{soc}
P. Bak, C. Tang, and K. Wiesenfeld, Phys. Rev. Lett. {\bf 59}, 381
(1987); Phys. Rev. A {\bf 38}, 364 (1988)
\bibitem{bak}
S. Maslov, M. Paczuski, and P. Bak, ``Avalanches and 1/f noise in evolution
and growth models", Phys. Rev. Lett. {\bf 73}, 2162 (1994)
\bibitem{teic}
S.B. Lowen and M.C. Teich, Phys. Rev. E {\bf 47}, 992 (1993)
\bibitem{p1}
W.H. Press, Commun. Astrophys. {\bf 7}, 103-119 (1978)
\bibitem{bd}
T.S. Bunch and P.C.W. Davies, Proc. R. Soc. A {\bf 360}, 117 (1978)
\bibitem{fp}
L.H. Ford and L. Parker, Phys. Rev. D {\bf 16}, 245 (1977); V. Sahni,
Class. Quantum Grav. {\bf 5}, L113 (1988) and references therein.
\bibitem{tw}
N.C. Tsamis and R.P. Woodard, ``The physical basis for infrared divergences
in inflationary quantum gravity", Class. Quantum Grav. {\bf 11}, 2969 (1994);
``Strong infrared effects in quantum gravity",
Ann. Phys. {\bf 238}, 1 (1995)
\bibitem{f2}
A. Folacci, Phys. Rev. D {\bf 46}, 2553 (1992)
\bibitem{hand}
P.H. Handel, Phys. Rev. Lett. {\bf 34}, 1492 (1975); Phys. Rev. A {\bf 22},
745 (1980)
\bibitem{hand1}
P.H. Handel, Phys. Rev. A {\bf 38}, 3082 (1988)
\bibitem{hand2}
P.H. Handel, in {\em Noise in Physical Systems and 1/f Noise}, Eds.
A. D'Amico and P. Mazzetti, (Elsevier, 1986); S. Weinberg, Phys. Rev.
{\bf 140B}, 516 (1965)
\bibitem{sok}
I. Yu. Sokolov, astro-ph/9405060 and 
Phys. Lett. A {\bf 223}, 163 (1996)


%%%%%%%%%%%%%%%%%%%%%%%%%%%%%%%%%%%%%%%%%%%%%%%%%%%%%%%%%%%%%%%%%%%%%%

%D. Boyanovsky, H.J. de Vega, and R. Holman, ``Noneq. evolution of scalar
%fields in FRW cosmologies", Phys. Rev. D {\bf 49}, 2769 (1994)
%- uses Schr\"odinger functional approach
%
%P. Elmfors, K. Enqvist, and I. Vilja, ``On the non-eq. early universe",
%Phys. Lett. B {\bf 326}, 37 (1994)
%
%J. Ellis and G. Steigman, ``Noneq. in the very early universe", Phys. Lett.
%B {\bf 89}, 186 (1980)
%
%B.L. Hu, ``Noneq. quantum fields in cosmology: comments on selected current
%topics", umdpp-95-051 (1994)
%
%A. Linde and A. Mezhlumian, ``Stationary universe", PL B {\bf 307}, 25 (1993)
%
%P.S. Wesson, ``Cosmological source of vacuum elm zero-point energy",
%Phys. Rev. A {\bf 44}, 3385 (1991) and subsequent papers by Puthoff and
%Santos
%%%%%%%%%%%%%%%%%%%%%%%%%%%%%%%%%%%%%%%%%%%%%%%%%%%%%%%%%%%%%%%%%%%%%%%%%%

\end{thebibliography}
\end{document}